\documentclass{osa-article}

\journal{oe}

\def\usv{U}
\articletype{Research Article}

\begin{document}

\title{Asymptotic dynamics of three-dimensional bipolar ultrashort
  electromagnetic pulses in an array of semiconductor carbon nanotubes}
\author{Eduard G. Fedorov,\authormark{1,2} Alexander V. Zhukov,\authormark{3,4} Roland Bouffanais,\authormark{3,*} Boris A. Malomed,\authormark{5,6} Herv{\'e} Leblond,\authormark{7} Dumitru Mihalache,\authormark{8,9} Nikolay N. Rosanov,\authormark{10,6,11} Mikhail B. Belonenko,\authormark{12,13,4} Thomas F. George\authormark{14}}

\address{
\authormark{1}Technion -- Israel Institute of Technology, Haifa 32000, Israel\\
\authormark{2}Vavilov State Optical Institute, 199053 Saint Petersburg, Russia\\
\authormark{3}Singapore University of Technology and Design, 8 Somapah Road, 487372 Singapore\\
\authormark{4}Entropique Group Ltd., 3 Spylaw Street, Maori Hill, 9010 Dunedin, New Zealand\\
\authormark{5}Department of Physical Electronics, School of Electrical Engineering, Faculty of Engineering, Tel Aviv University, 69978 Tel Aviv, Israel\\
\authormark{6}Saint Petersburg National Research University of Information Technologies, Mechanics and Optics (ITMO University), 197101 Saint Petersburg, Russia\\
\authormark{7}LUNAM Universit{\'e}, Universit{\'e}
  d'Angers, Laboratoire de Photonique d'Angers, EA 4464, 2 Boulevard
  Lavoisier, 49000 Angers, France\\
  \authormark{8}Academy of Romanian Scientists, 54
  Splaiul Independentei, Bucharest, RO-050094, Romania\\
  \authormark{9}Horia
  Hulubei National Institute of Physics and Nuclear Engineering, Magurele,
  RO-077125, Romania\\
  \authormark{10}Vavilov State Optical Institute,
  199053 Saint Petersburg, Russia\\
  \authormark{11}Ioffe Institute, Russian Academy of Sciences, 194021 Saint
  Petersburg, Russia\\
  \authormark{12}Laboratory of Nanotechnology,
  Volgograd Institute of Business, 400048 Volgograd, Russia\\
  \authormark{13}Volgograd State University, 400062 Volgograd, Russia\\
  \authormark{14}Office of the Chancellor, Departments of
  Chemistry \& Biochemistry and Physics \& Astronomy, University of
  Missouri-St. Louis, St. Louis, Missouri 63121, USA
}

\email{\authormark{*}bouffanais@sutd.edu.sg} 



\begin{abstract}
We study the propagation of three-dimensional bipolar ultrashort
  electromagnetic pulses in an array of semiconductor carbon nanotubes at
  times much longer than the pulse duration, yet still shorter than the
  relaxation time in the system. The interaction of the electromagnetic field
  with the electronic subsystem of the medium is described by means of
  Maxwell's equations, taking into account the field inhomogeneity along the
  nanotube axis beyond the approximation of slowly varying amplitudes and
  phases. A model is proposed for the analysis of the dynamics of an
  electromagnetic pulse in the form of an effective equation for the vector
  potential of the field. Our numerical analysis demonstrates the possibility
  of a satisfactory description of the evolution of the pulse field at large
  times by means of a three-dimensional generalization of the sine-Gordon and
  double sine-Gordon equations.
\end{abstract}

\section{Introduction}
Carbon nanotubes---quasi-one-dimensional macromolecules of carbon~\cite{1,2,3}
---are now considered to be one of the most promising materials with a high
potential of applicability in the development of the elemental base for modern
electronics. From the point of view of applications in optoelectronics, it is
of particular interest to study the properties of nanotubes with respect to
the peculiarity of their electronic structure.  The non-parabolicity of the
dispersion law for the conduction electrons of nanotubes~\cite{1,2,3}
(i.e. the energy dependency on the quasi-momentum) determines the essential
nonlinearity of their response to the action of the electromagnetic
field~\cite{4}. This circumstance provides the principal possibility of
observing various unique electromagnetic phenomena in structures based on
carbon nanotubes, including nonlinear diffraction and self-focusing of laser
beams~\cite{5,6}.

The successes of modern laser technologies in the field of the formation of
powerful electromagnetic radiation with given parameters, including extremely
short laser pulses with durations corresponding to several half-cycles of
field oscillations~\cite{7,8,9,10,11,12,13,14,15,16}, provides the impetus for
further comprehensive investigations in the field of light-matter interaction,
the results of which can subsequently form the basis for the newest elements
of modern electronics. In particular, one of such promising areas of research
is the study of the generation and propagation of laser beams and extremely
short spatio-temporal pulses in various media (e.g., see~\cite{17,18,19,20,21,22,23,24,25,26,27,28,29,29a,29b}) including arrays
of carbon nanotubes~\cite{30}.

The possibility of stable propagation of ultrashort electromagnetic pulses in
arrays of carbon nanotubes was first theoretically established in~\cite{30, 31} in a one-dimensional model, and was subsequently studied
in a two-dimensional model~\cite{32,33,34,35,36} in the framework of the
homogeneous field approximation along the axes of nanotubes. The approach used
in the above works did not take into account the effects associated with the
inhomogeneity of the field along the nanotube axes; therefore, in order to
obtain a more realistic description of field evolution, the mathematical model
was generalized to the case of two and three spatial dimensions, and was
supplemented by equations that take into account the redistribution of the
charge density in the system under the influence of the field inhomogeneity
along the axis of the nanotubes. As a result, the peculiarities of propagation
of solitary electromagnetic waves in arrays of carbon nanotubes were studied
both in two-dimensional~\cite{37,38} and three-dimensional~\cite{39,40,40a}
models, taking into account the effect of localization of the field along the
nanotube axes. In the course of numerical experiments carried out within the
framework of these studies, the possibility of stable propagation of bipolar
ultrashort electromagnetic pulses in the form of bipolar ``breather-like''
light bullets at distances significantly exceeding the characteristic
dimensions of the pulses along the direction of their motion was confirmed.

To date, a large amount of information has already been accumulated as a
result of extensive studies of various aspects of the interaction of extremely
short laser pulses with arrays of nanotubes, but there are still numerous
questions that require further clarification. In particular, from the point of
view of possible practical applications, it is instrumental to study the
evolution of the electromagnetic pulse field in an array of nanotubes at time
intervals that significantly exceed the characteristic pulse duration, but
that are still shorter than the relaxation time in the system. Such a problem
of considering the dynamics of a pulse at times several orders of magnitude
greater than its duration naturally arises in the case of the propagation of
extremely short pulses with a characteristic duration $\Delta t_{\rm{pulse}}
\sim 10^{-15}-10^{-14}$ s in the medium with the relaxation time
$t_{\rm{rel}}\sim 10^{-12}-10^{-11}$ s, which is quite achievable with modern
technologies for fabricating nanotube structures. A study of the asymptotic
dynamics of an ultimately short pulse at large times was carried out earlier
in~\cite{41} for a one-dimensional model in the homogeneous field
approximation along the nanotube axis. However, the results obtained within
that simplified framework cannot always be automatically extrapolated to the
behavior of a pulse in a model containing more than one spatial dimension.
Peculiarities of the behavior of a pulse in a non-one-dimensional geometry can
occur owing to the influence of the transverse effects associated with the
diffraction spreading of the wave packet. In this connection, it seems
expedient to solve the problem of modeling the evolution of parameters of an
extremely short laser pulse in an array of carbon semiconductor nanotubes,
taking into account the three-dimensional localization of the field.


\section{System configuration and key assumptions}


%

In this paper, we consider the propagation of a solitary electromagnetic wave
in the bulk array of single-walled semiconductor carbon nanotubes (CNTs)
embedded in a homogeneous dielectric medium. The nanotubes considered here are
of the ``zigzag" type $(m,0)$, where the number $m$ (not a multiple of three
for semiconductor nanotubes) determines the nanotube radius
$R=mb\sqrt{3}/2\pi$, and $b=1.42\times 10^{-8}\mathrm{~cm}$ is the distance
between neighboring carbon atoms~\cite{1,2,3}. The CNTs are supposed to be
arranged in such a way that their axes are parallel to the common $x$-axis,
and distances between adjacent nanotubes are much larger than their diameter,
so that we neglect the interaction between them. This latter assumption allows
us to consider the system as an electrically quasi-one-dimensional one, in
which electron tunneling between neighboring nanotubes can be neglected, and
electrical conductivity is possible only along the axis of nanotubes.

\begin{figure}[tbp]
    \centering
  \includegraphics[width=0.6\textwidth]{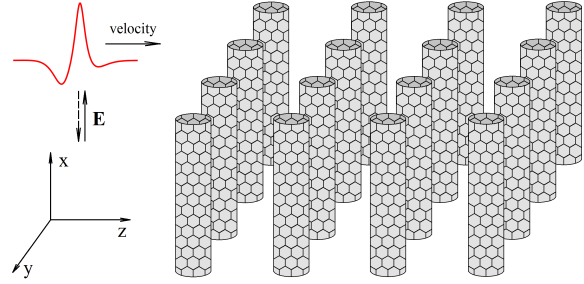}
  \caption{Schematic diagram of the considered system with the associated
    coordinate system.}
  \label{fig1}
\end{figure}

We define the configuration of the system in such a way that the pulse
propagates through the nanotube array in a direction perpendicular to their
axes (that is, for definiteness along the $z$-axis), and the electric
component of the wave field ${\bf E} = \{E,0,0\}$ is collinear with the
$x$-axis (see Fig.~\ref{fig1}).
We note that for a wide range of values of the system parameters, the
characteristic distance at which an appreciable change in the field of a
bipolar electromagnetic pulse occurs is significantly greater than the
distance between neighboring nanotubes, and also the characteristic length of
the conduction electron path along the axis of the nanotubes. For example, for
a nanotube radius $R\approx 2.74\times 10^{-8}\mathrm{~cm}$ and $m=7$, the
characteristic distance between them (sufficient to exclude the overlap of the
electron wave functions of neighboring nanotubes)---even if substantially
exceeding the value of the radius $R$---will be negligibly small in comparison
with the characteristic wavelength of the electromagnetic radiation in the
system.  In particular, in the case of infrared electromagnetic radiation this
assumption remains valid.  With this approximation, the nanotube array, which
is a discrete structure at the microscopic level, can nevertheless be
considered as a continuous medium at the considered wavelength scale of
electromagnetic radiation propagating in it.

Another important assumption that we adopt in this paper concerns the
magnitudes of the time duration of the electromagnetic pulse, $\Delta
t_{\rm{pulse}}$, the relaxation time of the conduction current along the
nanotube axis, $t_{\rm{rel}}$, and also the time interval for observing the
evolution of the field in the system, $\Delta t$. Specifically, we assume that
the observation time $\Delta t$ is much longer than the pulse duration $\Delta
t_{\rm{pulse}}$, but still shorter than the relaxation time $t_{\rm{rel}}$,
namely $\Delta t_{\rm{pulse}}\ll \Delta t < t_{\rm{rel}}$.

Given the orientation of the coordinate system axes relative to the nanotube
axis chosen above (see Fig.~\ref{fig1}), the electron energy spectrum for
carbon nanotubes takes the form
\begin{equation}
  \epsilon (p_{x},s)=\gamma_{0}\sqrt{1+4\cos \left( p_{x}\frac{d_{x}}{\hbar }%
    \right) \cos \left( \pi \frac{s}{m}\right) +4\cos ^{2}\left( \pi \frac{s}{m}%
    \right) },  \label{1}
\end{equation}
where the electron quasimomentum is $\mathbf{p}=\left\{ p_{x},s\right\} $, $s$
being an integer characterizing the momentum quantization along the perimeter
of the nanotube, $s=1,2,\dots ,m$, $m$ is the number of hexagonal carbon
cycles, forming the circumference of a nanotube, $\gamma_{0}$ is the overlap
integral, and where $d_{x}=3b/2$~\cite{1,2,3}.


\section{Evolution of the system parameters}


\subsection{Complete set of equations}



As we showed earlier (see, for example~\cite{40,40a} and references
therein), the evolution of the electromagnetic field in an array of nanotubes
is described by Maxwell's equations for the vector and scalar potentials,
${\bf A} = \{A,0,0\}$ and $\phi$, together with the continuity
equation~\cite{42, 43} for the current density ${\bf j} = \{j,0,0\}$, which is
calculated according to the approach proposed in~\cite{44,45}. As a
result, the set of evolutionary equations consists of three of them: for the
vector and scalar potentials, ${\bf A} = \{A,0,0\}$ and $\phi$, respectively,
and for the conduction electron density $n$.

The equation describing the evolution of the vector potential of a
self-consistent electromagnetic field in an array of nanotubes has the
form~\cite{40}

\begin{equation}
  \frac{\partial ^{2}\Psi }{\partial \tau ^{2}}-\left( \frac{\partial ^{2}\Psi}{\partial \xi ^{2}}+\frac{\partial ^{2}\Psi }
    {\partial \upsilon ^{2}}+\frac{\partial ^{2}\Psi }{\partial \zeta ^{2}}\right) +\eta \sum_{r=1}^{\infty
  }G_{r}\sin \left[ r\left( \Psi +\int\limits_{0}^{\tau }\frac{\partial \Phi }{\partial \xi }d\tau ^{\prime }\right) \right] =0.  
  \label{2}
\end{equation}
Here, we use the following notation: $\Psi =Aed_{x}/\left( c\hbar \right) $ is
the projection of the dimensionless vector potential onto the nanotube axis;
$\Phi =\phi\sqrt{\varepsilon }ed_{x}/(c\hbar )$ is the dimensionless scalar
potential; $\varepsilon$ is the average relative permittivity of the sample
(see, for example~\cite{46}); $\eta =n/n_{\rm{bias}}=\eta
(\xi,\upsilon,\zeta,\tau)$ is the reduced (dimensionless) distribution of
conduction electron density, with n the instantaneous value of the conduction
electrons concentration at an arbitrary point of the sample bulk;
$n_{\rm{bias}}$ is the concentration of conduction electrons in the sample in
the absence of an electromagnetic field (that is, essentially a constant
component of the concentration). The coefficients $G_r$ are the dimensionless
quantities that decrease with increasing $r$~\cite{40}; $\tau =\omega
_{0}t/\sqrt{\varepsilon }$ is the scaled time; $\xi =x\omega_{0}/c$, $\upsilon
=y\omega _{0}/c$ and $\zeta =z\omega _{0}/c$ are the dimensionless
coordinates. The quantity $\omega _{0}= 2 |e| d_{x}\hbar^{-1} \sqrt{\pi
  \gamma_{0}n_{\rm{bias}}}$ has the meaning of the characteristic frequency of
the electron subsystem of nanotubes ($e<0$ is the electron charge).

The equation describing the spatio-temporal evolution of the concentration of
conduction electrons in an array of nanotubes has the form~\cite{37,38,39,40,40a}
\begin{equation}
  \frac{\partial \eta }{\partial \tau }=\alpha \sum_{r=1}^{\infty }G_{r}\frac{%
    \partial }{\partial \xi }\left\{ \eta \sin \left[ r\left( \Psi
        +\int\limits_{0}^{\tau }\frac{\partial \Phi }{\partial \xi }d\tau ^{\prime
        }\right) \right] \right\} ,  \label{3}
\end{equation}
with $\alpha = d_{x}\gamma_{0}\sqrt{\varepsilon }/c\hbar $. This equation
determines the change in the electron concentration under the effect of a
self-consistent electromagnetic field in the array of nanotubes. We emphasize
that the inhomogeneity (localization) of the field only along the $x$-axis can
cause non-stationarity and dynamic spatial inhomogeneity of the electron
concentration in the sample, due to the presence of conductivity only along
the axis of the nanotubes. In this case, the inhomogeneity of the field along
the directions orthogonal to the axes of the nanotubes will not contribute to
the redistribution of the electron concentration due to the negligible overlap
of the electron wave functions of neighboring nanotubes and the absence of
conductivity in the $yOz$--plane.

As was shown earlier in~\cite{37,38,39,40,40a}, the equation describing the
evolution of the scalar potential of a field in an array of nanotubes has the
form

\begin{equation}
  \frac{\partial ^{2}\Phi }{\partial \tau ^{2}}-\left( \frac{\partial ^{2}\Phi}{\partial \xi ^{2}}
    +\frac{\partial ^{2}\Phi }{\partial \upsilon ^{2}}+\frac{\partial ^{2}\Phi }{\partial \zeta ^{2}}\right) 
  =\beta (\eta - \eta_0),  
  \label{4}
\end{equation}
where $\beta =1/\alpha=c\hbar /\left( d_{x}\gamma_{0}\sqrt{\varepsilon
  }\right)$, and $\eta_0 = n_0/n_{\rm{bias}} = \eta
(\xi,\upsilon,\zeta,\tau_0)$ is the reduced (dimensionless) value of the
concentration of conduction electrons at a given point in space at the initial
instant of time $\tau_0$ in the absence of the field.  In the simplest
particular case corresponding to an initially homogeneous sample ($n_0 =
n_{\rm{bias}}$), we have $\eta_0 = 1$.

Thus, the evolution of the field in the array of nanotubes is described by the
set of Eqs. \eqref{2}--\eqref{4}.  These equations describe a
self-consistent system of physical parameters, $\Psi$, $\Phi$, and $\eta$, the
dynamics of which reflect the mutual influence of the electromagnetic field
and the electronic subsystem of the nanotubes array (self-consistent
light-matter interaction).

\subsection{Effective equation for the vector potential}


When considering the process of propagation of an electromagnetic pulse in an
array of nanotubes at times considerably exceeding the pulse duration, special
attention should be given to an adequate consideration of the physical factors
affecting the evolution of the wave packet parameters. An analysis of the
degree of influence of these factors on the pulse dynamics in certain cases
can allow us to optimize the mathematical model describing the process. In
particular, one of such factors is the inhomogeneity of the electromagnetic
field along the nanotube axis. As follows from the set of Eqs.
\eqref{2}--\eqref{4}, the limiting (three-dimensional) localization of the
field in the array of nanotubes can cause a redistribution of the conduction
electron density. Thus, the propagation of the electromagnetic pulse occurs
when interacting with the dynamic inhomogeneities of the medium, which are
induced by the field of the pulse itself.

As shown by the numerical simulation performed earlier (see~\cite{37,
  39, 40}), the electron density differences (dynamic inhomogeneities) formed
during the passage of electromagnetic pulses in the sample have magnitude of
the order of one percent relative to the initial equilibrium concentration
$n_0$.  In this case, there was no dramatic change in the dynamics of the
pulses with respect to the results obtained within the framework of the model
limited to the approximation of a homogeneous field along the nanotube axis
(see, for example~\cite{32}). This can be explained by the fact that the
higher the speed of the electromagnetic pulse, the shorter the time during
which the pulse affects the electronic subsystem of the array of nanotubes,
the less is the formation of dynamic electron density inhomogeneities.  At the
same time, the higher the pulse speed, and the smaller the time during which
the pulse propagates in the region of the space containing the dynamic
inhomogeneities of the electron concentration, which cause the pulse field to
be adjusted to the changed properties of the medium. Thus, under the
conditions considered in this paper, the time of interaction between the field
and the perturbations of the medium is not sufficient to generate a noticeable
change in the shape of the pulse. Thus, in the case of ultrashort
electromagnetic pulses (when the condition $\Delta t_{\rm{pulse}} \ll
t_{\rm{rel}}$ is satisfied), the non-stationarity of the conduction electron
density distribution can be neglected. Proceeding from these considerations,
we will assume that the distribution of the conduction electron concentration
in the sample remains approximately constant, that is, $\partial\eta
/\partial\tau = 0$. If the concentration of conduction electrons in the
nanotubes array initially is uniform throughout the sample volume, i.e., there
were no regions of high or low electron concentration (that is,
$n_0=n^{\rm{bias}}$), then it can be assumed that $\eta \approx \eta_0 = 1$
throughout the sample during the entire observation time $t <
t_{\rm{rel}}$. Under the condition $\eta = \eta_0$, the right-hand side of
Eq.~\eqref{4} vanishes, and without loss of generality (see~\cite{42, 43}), we can safely impose that
$\Phi(\xi,\upsilon,\zeta,\tau) = 0$.

Thus, taking into account the above arguments, one can exclude Eqs.~\eqref{3}
and \eqref{4} for the concentration of conduction electrons and the scalar
potential from further consideration. As a result, taking into account the
assumptions made about the short duration of the electromagnetic pulse
propagating in the array of nanotubes, with sufficient accuracy, the evolution
of the field in a homogeneous array of nanotubes (which does not initially
contain regions of increased or decreased electron concentration) can be
described by the only equation resulting from Eq.~\eqref{2} with $\eta =1$ and
$\Phi =0$:
\begin{equation}
  \frac{\partial ^{2}\Psi }{\partial \tau ^{2}}-\left( \frac{\partial ^{2}\Psi}{\partial \xi ^{2}}+\frac{\partial ^{2}\Psi }
    {\partial \upsilon ^{2}}+\frac{\partial ^{2}\Psi }{\partial \zeta ^{2}}\right) + \sum_{r=1}^{\infty
  }G_{r}\sin \left( r\Psi\right) =0.  
  \label{5}
\end{equation}
This equation is an effective equation describing the evolution of the field
in an array of nanotubes in the case of propagation of ``fast"
three-dimensional extremely short electromagnetic pulses. Since Eq.~\eqref{5}
contains all the terms of the series $G_r$---where the subscript $r$
corresponds to the mode number in the expansion of the conduction electron
energy~\eqref{1} in the Fourier series (see, for example, Eq.~(4) in~\cite{40}), this equation can be regarded as the ``complete" effective
equation.

Calculations show that the coefficients $G_r$ decrease rapidly with increasing
values of $r$ (see~\cite{40} and references therein). Therefore, in a
number of cases, for a qualitative approximate description of the evolution of
the electromagnetic field in the system under consideration, we can restrict
ourselves to only two terms with $r=\{1,2\}$ under the summation sign in
Eq.~\eqref{5}, thereby leading to
\begin{equation}
  \frac{\partial ^{2}\Psi }{\partial \tau ^{2}}-\left( \frac{\partial ^{2}\Psi}{\partial \xi ^{2}}+\frac{\partial ^{2}\Psi }
    {\partial \upsilon ^{2}}+\frac{\partial ^{2}\Psi }{\partial \zeta ^{2}}\right) + G_{1}\sin \left( \Psi\right) +
  G_{2}\sin \left( 2\Psi\right) =0.  
  \label{6}
\end{equation}
Equation \eqref{6} can be regarded as a ``reduced" effective equation, which
happens to be a three-dimensional generalization of the Double sine-Gordon
equation~\cite{47}. To the best of our knowledge, this represents the first
attempt in using the sine-Gordon equations for the three-dimensional study of
such pulse propagation in arrays of CNTs. It is worth adding that using model~\eqref{6}, as opposed to model~\eqref{5}, provides significant savings from the computational standpoint.

The case of the most radical reduction of the total effective Eq.~\eqref{5}
deserves special consideration. In the case of a significant excess of the
absolute value of the quantity $G_1$ over the absolute value of the quantity
$G_2$, we can restrict the summation to the first term under the summation
sign, having obtained a non-one-dimensional generalization of the sine-Gordon
equation:
\begin{equation}
  \frac{\partial ^{2}\Psi }{\partial \tau ^{2}}-\left( \frac{\partial ^{2}\Psi}{\partial \xi ^{2}}+\frac{\partial ^{2}\Psi }
    {\partial \upsilon ^{2}}+\frac{\partial ^{2}\Psi }{\partial \zeta ^{2}}\right) + \sigma^2\sin \left( \Psi\right) =0,  
  \label{7}
\end{equation}
where $\sigma = \sqrt{G_1}$ ($G_1>0$). It is worth adding that in the present
case, the amplitude of the 3D breathers is not small. Thus, linearizing the
sine-Gordon equations is not an option. However, a Taylor expansion with
respect to $\Psi$, up to quintic terms, could be considered. Doing so would
enable the use 3D nonlinear equations with cubic-quintic nonlinear terms, in
which the classical variational approximation could be applied to 3D
breathers. Note that these equations with the cubic-quintic nonlinearity are
known to support stable solitons with embedded vorticity~\cite{47b}.

Following the reasoning proposed in~\cite{41}, the reduced effective
equation for the vector potential \eqref{7} can be replaced by an equation of
the form
\begin{equation}
  \frac{\partial ^{2}\Psi }{\partial \tau ^{2}}-\left( \frac{\partial ^{2}\Psi}{\partial \xi ^{2}}+\frac{\partial ^{2}\Psi }
    {\partial \upsilon ^{2}}+\frac{\partial ^{2}\Psi }{\partial \zeta ^{2}}\right) + 
  \sigma_\Sigma^2\sin \left( \Psi\right) =0,  
  \label{8}
\end{equation}
where $\sigma_\Sigma$ is determined by the empirical formula~\cite{41}
\begin{equation}
  \sigma_\Sigma = \sqrt{\sum_{r=1}^\infty G_r}.
  \label{9}
\end{equation}
The model simplification in the form of the 3D generalization of the sine-Gordon equation conveniently offers the ability to estimate analytically the asympotic dynamics of the electromagnetic pulses (see Appendix). The complete effective Eq.~\eqref{5} and the three versions of the
reduced equation, \eqref{6}--\eqref{8}, will be used below as comparative
models for studying the dynamics of the ultrashort electromagnetic pulse in an
array of semiconductor carbon nanotubes over long times. Laslty, it is worth
adding that our effective simplified model in the form of Eq.~\eqref{5} (or
alternatively in the form of Eqs.~\eqref{6}--\eqref{8}) is primarily
applicable to the description of the central part of the pulse, rather than
its tail.


\section{Characteristics of the electromagnetic pulse field}

As a quantity illustrating the localization of an electromagnetic pulse in
space, we will use the energy characteristic of the field, which is
proportional to the bulk density of the field energy. For such a value, it is
convenient to take the square of the electric field strength along the axis of
the nanotubes, $I(\xi,\upsilon,\zeta,\tau) = E_x^2$. This quantity will be
referred to as ``intensity" for convenience.

Using the well-known formula $\mathbf{E} = -c^{-1}\partial\mathbf{A}/\partial
t - \nabla\phi$ (see, e.g.~\cite{42, 43}), and also taking into account
that the vector potential has a nonzero component only along the nanotube axis
(see the description of the system configuration above), one can simply
express the electric field vector $\mathbf{E} = \{E,0,0\}$ as

\begin{equation}
  E_x=E_{0}\frac{\partial \Psi }{\partial \tau } ,  
  \label{10}
\end{equation}
where $E_{0}= -\hbar \omega _{0}/\sqrt{ed_{x}\varepsilon }$. Taking into
account Eq.~\eqref{10}, the energy characteristic of the field chosen above
can be represented as
\begin{equation}
  I=I_{0}\left( \frac{\partial \Psi }{\partial \tau }\right) ^{2},  \label{11}
\end{equation}
where $I_{0}=E_0^2$.

To illustrate the position of the electromagnetic pulse in space at an
arbitrary time instant $\tau$, we define the dimensionless averaged
characteristic coordinates of the pulse as the coordinates of the pulse
centroid (``center of mass"), $\left\{
  \xi_{\rm{pulse}},\upsilon_{\rm{pulse}},\zeta_{\rm{pulse}}\right\}$.  To this
end we calculate the first order of the moments of the pulse intensity,
following the approach proposed in~\cite{48}:
\begin{equation}
  \xi_{\rm{pulse}} (\tau) = \frac{\int_{-\infty}^\infty \int_{-\infty}^\infty \int_{-\infty}^\infty
    \xi I(\xi,\upsilon,\zeta,\tau) d\xi d\upsilon d\zeta} {\int_{-\infty}^\infty \int_{-\infty}^\infty 
    \int_{-\infty}^\infty I(\xi,\upsilon,\zeta,\tau) d\xi d\upsilon d\zeta},
  \label{12}
\end{equation}

\begin{equation}
  \upsilon_{\rm{pulse}} (\tau) = \frac{\int_{-\infty}^\infty \int_{-\infty}^\infty \int_{-\infty}^\infty
    \upsilon I(\xi,\upsilon,\zeta,\tau) d\xi d\upsilon d\zeta} {\int_{-\infty}^\infty \int_{-\infty}^\infty 
    \int_{-\infty}^\infty I(\xi,\upsilon,\zeta,\tau) d\xi d\upsilon d\zeta},
  \label{13}
\end{equation}

\begin{equation}
  \zeta_{\rm{pulse}} (\tau) = \frac{\int_{-\infty}^\infty \int_{-\infty}^\infty \int_{-\infty}^\infty
    \zeta I(\xi,\upsilon,\zeta,\tau) d\xi d\upsilon d\zeta} {\int_{-\infty}^\infty \int_{-\infty}^\infty 
    \int_{-\infty}^\infty I(\xi,\upsilon,\zeta,\tau) d\xi d\upsilon d\zeta}.
  \label{14}
\end{equation}

The speed of the electromagnetic pulse is then defined as
\begin{equation}
  {\bf v} = \frac{c}{\sqrt{\varepsilon}} \left\{ \frac{d\xi_{\rm{pulse}}}{d\tau}, 
    \frac{d\upsilon_{\rm{pulse}}}{d\tau}, \frac{d\zeta_{\rm{pulse}}}{d\tau} \right\}.
  \label{15}
\end{equation}

As quantitative characteristics of the localization of the pulse field in
space, we calculate the longitudinal half-width of the pulse $\lambda_\zeta$
(along the $\zeta$-axis), as well as the transverse half-widths $\lambda_\xi$
and $\lambda_\upsilon$ along the $\xi$-axis and $\upsilon$-axis,
correspondingly. We employ the M-squared method for characterizing laser
beams~\cite{48}. We adopt this approach to the situation corresponding to the
limiting (three-dimensional) localization of the field, therefore defining the
half-widths of the pulse by means of the second moment of the pulse intensity
profile as follows:
\begin{equation}
  \lambda_\xi (\tau) = \frac{\int_{-\infty}^\infty \int_{-\infty}^\infty \int_{-\infty}^\infty
    \left( \xi - \xi_{\rm{pulse}} (\tau)\right)^2 I(\xi,\upsilon,\zeta,\tau) d\xi d\upsilon d\zeta} {\int_{-\infty}^\infty \int_{-\infty}^\infty 
    \int_{-\infty}^\infty I(\xi,\upsilon,\zeta,\tau) d\xi d\upsilon d\zeta},
  \label{16}
\end{equation}
\begin{equation}
  \lambda_\upsilon (\tau) = \frac{\int_{-\infty}^\infty \int_{-\infty}^\infty \int_{-\infty}^\infty
    \left( \upsilon - \upsilon_{\rm{pulse}} (\tau)\right)^2 I(\xi,\upsilon,\zeta,\tau) d\xi d\upsilon d\zeta} {\int_{-\infty}^\infty \int_{-\infty}^\infty 
    \int_{-\infty}^\infty I(\xi,\upsilon,\zeta,\tau) d\xi d\upsilon d\zeta},
  \label{17}
\end{equation}
\begin{equation}
  \lambda_\zeta (\tau) = \frac{\int_{-\infty}^\infty \int_{-\infty}^\infty \int_{-\infty}^\infty
    \left( \zeta - \zeta_{\rm{pulse}} (\tau)\right)^2 I(\xi,\upsilon,\zeta,\tau) d\xi d\upsilon d\zeta} {\int_{-\infty}^\infty \int_{-\infty}^\infty 
    \int_{-\infty}^\infty I(\xi,\upsilon,\zeta,\tau) d\xi d\upsilon d\zeta}.
  \label{18}
\end{equation}
For simplicity, we introduce the generalized transverse characteristic
half-width of the pulse, $\lambda_\perp$, as:
\begin{equation}
  \lambda_\perp = \frac{1}{2} \left( \lambda_\xi (\tau) + \lambda_\upsilon (\tau)\right).
  \label{19}
\end{equation}


\section{Numerical experiment and discussion of the results}


\subsection{Initial condition: shape of the electromagnetic pulse}
We will investigate the evolution of the field of an electromagnetic pulse in
an array of nanotubes numerically, solving the equation for the vector
potential (see Eqs. \eqref{5}--\eqref{8}).  As an initial condition, we
take an ``instantaneous snapshot" of the distribution of the dimensionless
projection of the field vector potential on the $\xi$-axis (the axis of the
nanotubes) at the time instant $\tau = \tau_0$:
\begin{equation}
  \Psi(\xi,\upsilon,\zeta,\tau_0) = \Psi_\| (\zeta) \Psi_\perp (\xi, \upsilon),
  \label{20}
\end{equation}
where the function $\Psi_\| (\zeta,\tau_0)$ determines the distribution (along the $Oz$ direction) of the projection of the dimensionless vector potential on the nanotube axis $(Ox)$; and the function
$\Psi_\perp (\xi, \upsilon)$ represents the initial distribution of the field
in the $\xi O\upsilon$--plane orthogonal to the direction of propagation of
the pulse.

We select the profile $\Psi_\| (\zeta,\tau_0)$ corresponding to the breather
solution of the sine-Gordon equation, namely, to the non-topological
oscillating soliton~\cite{47},
\begin{equation}
  \Psi_\| (\zeta,\tau_0) = 4\arctan \left\{ \left(\frac{1}{\Omega ^{2}}-1\right)^{1/2}
    \frac{\sin \chi}{\cosh \mu}\right\} ,  
  \label{21}
\end{equation}
where
\begin{align}
  \chi & = \sigma \Omega\frac{\tau _{0}-(\zeta -\zeta_{0})\usv }{\sqrt{1-\usv ^2}},  \label{22} \\
  \mu & = \sigma \left[ \tau _{0}\usv -(\zeta -\zeta_{0}) \right]
  \sqrt{\frac{1-\Omega^{2}} {1-\usv^2}},
  \label{23}
\end{align}
with $\usv=u/v_0$ being the ratio between the initial propagation velocity $u$
of the sine-Gordon breather in the one-dimensional approximation along the
$\zeta$-axis, and the speed of light in the medium $v_0 =
c/\sqrt{\varepsilon}$. The quantity $\zeta_0$ is the dimensionless coordinate
of the center of mass of breather~\eqref{21} along the $\zeta$-axis at the
time instant $\tau = \tau_0$, and $\Omega = \omega_B/\omega_0$, where
$\omega_B$ is the self-oscillation frequency of the breather ($0<\Omega <
1$). As for the value of the quantity $\sigma$ in Eqs.~\eqref{22} and
\eqref{23}, we choose $\sigma = \sqrt{G_1}$ for certainty and comparability of
the results of modeling the evolution of the same initial pulse in different
models (see Eqs.~\eqref{5}--\eqref{8}); the values of the
coefficients $G_r$ are calculated using Eq.~(4) from~\cite{40}.  The
justification of the choice of the initial pulse profile~\eqref{21} is given
in our previous papers (e.g., see~\cite{38,40}).

We choose the transverse profile of the intensity of the pulse field with the
corresponding Gaussian distribution, which is justified by the high degree of
adequacy of this description to real situations widely represented in various
fields of physics and technology~\cite{5,6,49,50,51}:
\begin{equation}
  \Psi_\perp (\xi,\upsilon) = \exp\left\{ -\frac{(\xi-\xi_0)^2+(\upsilon-\upsilon_0)^2}{w_0^2}\right\},  
  \label{24}
\end{equation}
where $\xi_0$ and $\upsilon_0$ are the dimensionless coordinates of the
``center of mass" of the pulse at the initial instant of time $\tau_0$, and
$w_0$ is the value characterizing the transverse localization of the
electromagnetic pulse at the initial instant of time (the value of this
characteristic is proportional to the initial value $\lambda_\perp (0)$
determined by Eq.~\eqref{19}, but not exactly equal to it).

Taking into account the expressions \eqref{20}--\eqref{24}, the projection of
the electric field strength \eqref{10} of the electromagnetic pulse field onto
the axis of nanotubes at the initial instant of time has the form
\begin{equation}
  E_x = E_{\rm{max}}\frac{\cos\chi \cosh\mu - \usv \left( \Omega^{-2} - 1\right)^{1/2}
    \sin\chi \sinh\mu}{\cosh^2\mu +\left( \Omega^{-2} - 1\right)\sin^2\chi}
  \exp\left\{ -\frac{(\xi-\xi_0)^2+(\upsilon-\upsilon_0)^2}{w_0^2}\right\} ,
  \label{25}
\end{equation}
where $E_{\rm{max}} = 4E_0\sigma\sqrt{1-\Omega^2}/\sqrt{1-\usv^2}$.

The electromagnetic pulse corresponding to the breathing solution of the form
\eqref{21} of the sine-Gordon equation for the vector potential, during the
propagation, performs internal oscillations: the shape of its profile. The projection of the electric field
strength of such a pulse takes both positive and negative values, so a pulse
of this type is said to be ``bipolar."

The duration of the electromagnetic pulse with a profile whose shape can be
approximately described by Eq.~\eqref{21} is determined by the function
$\cosh\mu$ that determines the envelope of a given wave packet. The duration
of the electromagnetic pulse $\Delta t_{\rm{pulse}}$ is defined as the time
interval during which the values of the ``running" envelope (i.e., the
function $\cosh\mu$) measured at a fixed point will exceed half of its peak
amplitude value (i.e., have values above the level $0.5$).  Given that the
width of the function $\cosh\mu$ at the level of $0.5$ (FWHM -- full width of
half maximum) is $\Delta\mu_{0.5} = 2\ln \left( 2+\sqrt{3}\right)$, we get the
value of the duration of the electromagnetic pulse:
\begin{equation}
  \Delta t_{\rm{pulse}} = 2 \frac{\ln \left( 2+\sqrt{3}\right)}{\sigma\omega_0} \frac{\sqrt{\varepsilon}}{\usv}
  \frac{\sqrt{1-\usv^2}}{\sqrt{1-\Omega^2}}.
  \label{26}
\end{equation}


\subsection{System parameters}

As an environment for modeling the propagation of an extremely short
electromagnetic pulse, we chose an array of nanotubes of the zig-zag type
$(m,0)$: $m=7$, $\gamma_0 = 2.7$ eV, $b=1.42\times 10^{-8}$ cm, $d_x\approx
2.13 \times 10^{-8}$ cm, $n_{\rm{bias}} = 10^{18}$ cm$^{-3}$ at temperature
$T=293$ K. We assume that the array of CNTs is embedded in a dielectric matrix
with the effective dielectric constant $\varepsilon = 4$.

The dimensionless parameter $\usv$ (see Eqs.~\eqref{21}--\eqref{23}) was
varied within the interval $\usv \in (0.5;0.999)$.  For $\usv < 0.5$ and
further decrease in the value of this parameter, the longitudinal width of the
electromagnetic pulse begins to approach the value of the distance traveled by
the pulse over a duration $\sim t_{\rm{rel}}$, which is of no significant
practical interest.  The case corresponding to the condition $\usv > 0.999$ is
not described in the present paper because of the limitations of the numerical
scheme we used.

The dimensionless parameter of the frequency of the internal oscillations
$\Omega$ of the initial pulse~\eqref{21} was varied over the interval
$\Omega\in (0.1;0.9)$. As this parameter decreases, the characteristic width
of the pulse along the $\zeta$-axis decreases, while for $\Omega<0.5$, the
variation in the form of the profile is insignificant. For $\Omega > 0.9$, the
width of the pulse along the $\zeta$-axis becomes comparable with the
dimensions of the numerical grid.  The parameter $w_0$ of the characteristic
transverse width of the pulse was varied in the range from $1.0$ to $10.0$.

The set of differential Eqs.~\eqref{5}--\eqref{8} does not have an exact
analytical solution in the general case. Therefore, we carried out numerical
simulations to study the propagation of an electromagnetic pulse in an array
of CNTs. To solve each of the equations with initial
conditions~\eqref{20}--\eqref{25}, we used an explicit finite-difference
three-layered scheme of the cross type described in~\cite{52,53,54} and
adapted by us for the three-dimensional model, using the approach developed
and detailed in~\cite{38}. As a result of the numerical experiment, we
have modeled the evolution of the vector potential
$\Psi(\xi,\upsilon,\zeta,\tau)$, and also calculated the distribution of the
field strength and intensity using Eqs.~\eqref{10} and~\eqref{11}.

We emphasize that the mathematical model used in this work is valid for
observation times $t$ shorter than the relaxation time $t_{\rm{rel}}$ of the
electronic subsystem, since under the condition $t<t_{\rm{rel}}$, only the
evolution of the electromagnetic field in the system can be adequately
evaluated, neglecting damping due to collisions of electrons with
irregularities in the crystal structure of the nanotube array. For example,
with $t_{\rm{rel}}\approx 10^{-11}$ s---which can be achieved with modern
sampling technologies---the limiting distance traveled by light in the medium
under consideration is $\Delta z \approx
ct_{\rm{rel}}/\sqrt{\varepsilon}=0.15$ cm.

We performed a numerical experiment on the propagation of an electromagnetic
pulse over the time interval $\Delta\tau = 3\times 10^2$ (corresponding to
dimensional time $\Delta t = \Delta\tau \sqrt{\varepsilon}/\omega_0 \approx
8.4\times 10^{-12}$ s), which is close in the order of magnitude to the value
of the relaxation time $t_{\rm{rel}}$ indicated above. The purpose of the
numerical experiment was to find out whether the solutions of equations
~\eqref{5}--\eqref{8} retain the properties inherent to the initial condition
in the form of a three-dimensional pulse with a longitudinal profile of the
breather (see~\eqref{21}). Specifically, we were interested in whether the
individuality of the electromagnetic pulse persists in spite of the phenomena
of dispersive and diffractional spreading, whether the wave packet remains
bipolar, and whether it retains the properties of the breather with respect to
periodic changes in the shape and amplitude over long times.

For definiteness and comparability, the results of modeling the propagation of
an electromagnetic pulse in an array of carbon nanotubes is presented below
for the following values of the dimensionless parameters of the model: $\Omega
= 0.5$, $w_0 = 5.0$, $\usv = 0.95$ ($u = 1.425\times 10^{10}$ cm/s). The
maximum amplitude of the electric field of the pulse in this case has a value
$| E_x |_{\rm{max}} \approx 1.170\times 10^7$ V/cm (see Eq.~\eqref{25}), and
its characteristic duration is $\Delta t_{\rm{pulse}} \approx 2.9\times
10^{-14}$~s (see Eq.~\eqref{26}).  The characteristic frequency of the
electromagnetic field of a given sub-cycle pulse can be estimated as
$\omega_{\rm{pulse}} \approx 2\pi/\Delta t_{\rm{pulse}}\approx 2.14\times
10^{14}$ s$^{-1}$.

The results of the comparison of the numerical solution of the full effective
Eq.~\eqref{5} with the solutions of the reduced Eqs.~\eqref{6}--\eqref{8}, with the same initial
condition~\eqref{20}--\eqref{25} for all these four cases, are given below.


\subsection{The complete effective equation}

\begin{figure}[tbp]
    \centering
  \includegraphics[width=1.0\textwidth]{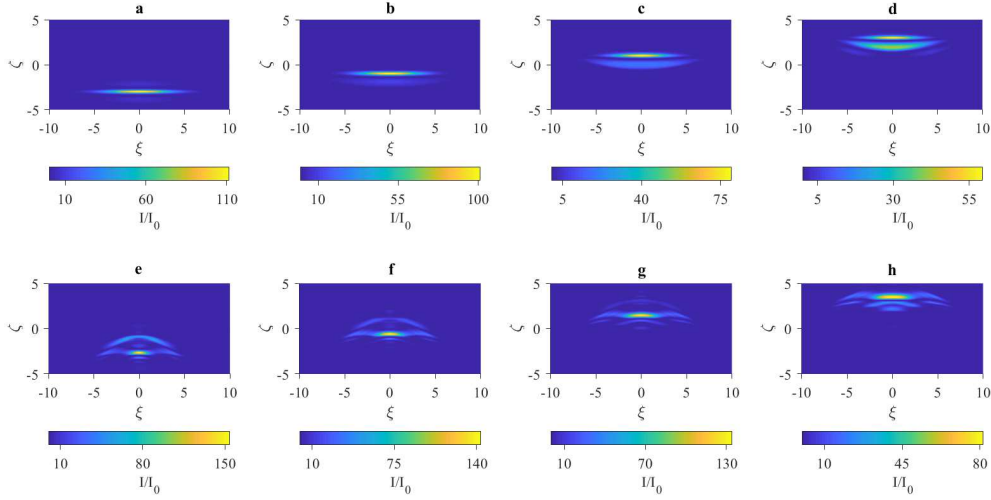}
  \caption{Distribution of the quantity $I/I_0 = \left( \partial\Psi
      / \partial\tau \right)^2$ in an array of nanotubes at various instants
    of dimensionless time $\tau$ during the passage of the wave packet, which
    is the solution of Eq.~\eqref{5}: (a) $\tau = 0$, (b) $\tau = 2.0$, (c)
    $\tau = 4.0$, (d) $\tau = 6.0$, (e) $\tau = 28.0$, (f) $\tau = 30.0$, (g)
    $\tau = 32.0$, (h) $\tau = 34.0$.}
  \label{fig2}
\end{figure}

\begin{figure}[tbp]
\centering
  \includegraphics[width=1.0\textwidth]{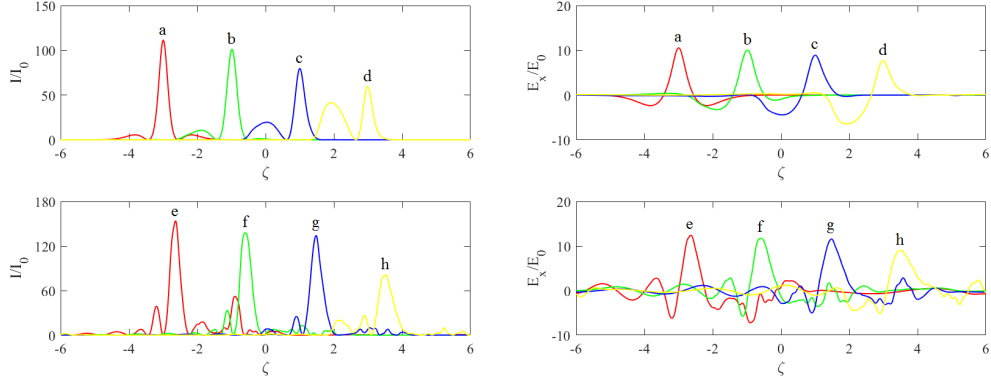}
  \caption{Distribution of the electric field strength $E_x/E_0$ and energy
    intensity $I/I_0$ on the $\zeta$-axis, in the course of the propagation of
    the electromagnetic pulse shown in Fig.~\ref{fig2}, at various instants of
    dimensionless time: (a) $\tau = 0$, (b) $\tau = 2.0$, (c) $\tau = 4.0$,
    (d) $\tau = 6.0$, (e) $\tau = 28.0$, (f) $\tau = 30.0$, (g) $\tau = 32.0$,
    (h) $\tau = 34.0$. }
  \label{fig3}
\end{figure}

\begin{figure}[tbp]
\centering
  \includegraphics[width=0.6\textwidth]{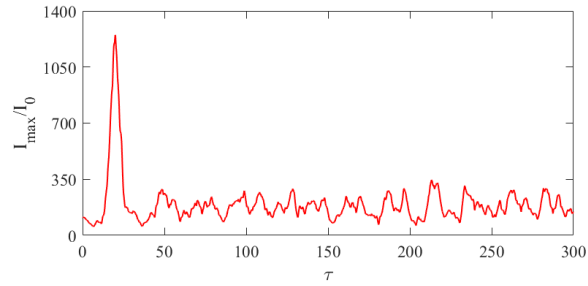}
  \caption{Dependency of the ratio $I_{\rm{max}}/I_0$ on the dimensionless
    time $\tau$ during the propagation of the electromagnetic pulse shown in
    Figs.~\ref{fig2} and~\ref{fig3}.  The dimensionless time $\tau = \omega_0
    t/\sqrt{\varepsilon}$ is plotted along the horizontal axis. }
  \label{fig4}
\end{figure}

Figures \ref{fig2}--\ref{fig4} show the results of the numerical simulations
of the propagation of a three-dimensional electromagnetic pulse within the
framework of the model represented by the full effective Eq.~\eqref{5}. Figure \ref{fig2} shows the distribution of the intensity
of the field, $I(\xi,\upsilon_0,\zeta,\tau) = I_0 \left( \partial\Psi
  / \partial\tau \right)^2$, in the $\xi O \zeta$ cross section (at $\upsilon
= \upsilon_0$) at various instances of the dimensionless time $\tau$.
The color scale is assigned to different values of the ratio $I/I_0$: the
yellow areas correspond to the maximum values, and the dark-blue areas
correspond to the minimum values. The horizontal and vertical axes of the
graph represent the dimensionless coordinates $\xi = x\omega_0/c$ and $\zeta =
z\omega_0/c$. Given the system parameter values selected above, the unit along
the $\xi$- and $\zeta$-axes corresponds to the distance $\approx 4.2\times
10^{-4}$~cm. We note that here we give the distribution of the quantity
$I/I_0$ in the $\xi O\zeta$--plane only, since the pattern of the distribution
of the intensity of the field in the $\upsilon O\zeta$--plane is qualitatively
similar.

We draw attention to the fact that in order to save computation time and to
simplify the visual representation, we used a numerical scheme with periodic
boundary conditions (see the details in our previous work~\cite{38}). For this
reason, in Figure~\ref{fig2} (and also in Figure~\ref{fig3}), the positions of
the wave packet at the later instants (``e''-- ``h'') belong to the same
spatial interval along the $\zeta$-axis that corresponds to the positions of
the wave packet at the earlier time instants (``a''--``d'').  It can be seen
from Fig.~\ref{fig2} that the wave packet, having overcome a distance
substantially exceeding its characteristic size along the direction of
propagation (along the positive $\zeta$ direction), retains its individuality
without undergoing a decay due to diffraction and dispersive spreading.

Figure~\ref{fig3} shows the evolution of the distribution profile of the
electric component of the field $E_x$~(Eq.~\eqref{10}) of a given wave packet
and the intensity $I$ along the $\zeta$-axis passing through the point with
coordinates $\xi = \xi_0$ and $\upsilon = \upsilon_0$, corresponding to the
initial position of the ``center of mass" of the pulse in the $\xi
O\zeta$--plane at the same instants of dimensionless time $\tau$ as in
Fig.~\ref{fig1}. In this case, the center of mass of the pulse in the $\xi
O\zeta$--plane is not displaced, i.e. according to the simulation results,
$\xi_{\rm{pulse}} = \xi_0$ and $\upsilon_{\rm{pulse}} = \upsilon_0$.

It can be seen from Fig.~\ref{fig3} that during the observation time, $E_x$ is
alternating between positive and negative values, that is, the electromagnetic
pulse remains bipolar. The nature of the change in the configuration of the wave packet can be
illustrated by the dependency of the scaled amplitude of the intensity
$I_{\rm{max}}/I_0 = {\rm{max}} \left\{ (\partial\Psi/\partial\tau)^2\right\}$
on the dimensionless time $\tau$. Figure~\ref{fig4} shows the time dependency
of the quantity $I_{\rm{max}}/I_0$ for the same values of the system
parameters as in Figs.~\ref{fig2} and~\ref{fig3}.  

As a result of numerical analysis, it is established that the wave packet
preserves its individuality in the course of propagation, without undergoing
significant spreading and damping; the electric field remains alternating over a time interval
substantially exceeding the characteristic duration of the wave packet. Thus,
this three-dimensional electromagnetic pulse has the properties of a
nontopological soliton--breather, determined by the initial
conditions~\eqref{21} and~\eqref{25}. This circumstance allows one, in a
certain sense, to consider a three-dimensional bipolar electromagnetic
extremely short pulse propagating in an array of carbon nanotubes as being a
soliton.

We note that the results presented here correspond to the situation in which
the electromagnetic pulse propagates in an essentially nonlinear regime. We
emphasize that in this case, the absolute values of the extrema of the
projection of the dimensionless vector potential $\Psi$ on the nanotubes axis
reach values of the order of unity (that is, the condition $|\sin\Psi | \ll 1$
is not satisfied in the general case) and, consequently, Eq.~\eqref{5} cannot
be linearized with the chosen values of the system parameters. From the
physical point of view, this means that the response $j$ (electric current
density) depends in an essentially nonlinear manner on the vector potential
(and hence on the electric field strength) of the self-consistent field of the
electromagnetic pulse (see formula (3) from our previous work~\cite{40}).


\subsection{Reduced effective equation}


\subsubsection{3D generalization of the double sine-Gordon equation}

As noted above, the coefficients $G_r$ (see Eq.~\eqref{5}) decrease rapidly
with increasing values of the index $r$. According to Eq.~(4) from our
previous work~\cite{40}, and with the system parameters chosen above, we
obtain the following values for the first five terms of the series $G_r$ (up
to the second decimal place): $G_1 \approx 0.91$, $G_2 \approx 0.33$, $G_3
\approx 0.18$, $G_4 \approx 0.12$, and $G_5 \approx 0.08$. Thus, in fact, the
third term is already significantly smaller than the first one, which
justifies the actual truncation to obtain the full effective equation, so that
Eq.~\eqref{6} is a satisfactory approximation of the original Eq.~\eqref{5}.

We have numerically solved the truncated effective Eq.~\eqref{6} in the form
of a three-dimensional generalized double sine-Gordon equation with initial
conditions~\eqref{20}--\eqref{25} on the dimensionless time interval ranging
from $0$ to $\tau = 300$. As a result, we established that the electromagnetic
pulse propagates over a distance substantially exceeding its characteristic
size, and does not undergo appreciable diffraction, transverse or dispersive
longitudinal spreading.  At the same time, the solution of the truncated
effective Eq.~\eqref{6} is a bipolar solitary electromagnetic wave with a
``breathing" periodically repeating its shape profile. In other words, the
evolution of the pulse field in model~\eqref{6} is qualitatively similar to
the behavior of the electromagnetic pulse in the model of the full effective Eq.~\eqref{5}. Thus, the evolution of the field of a three-dimensional
electromagnetic pulse (with the same initial parameters) in an array of
nanotubes is described by Eqs.~\eqref{5} and ~\eqref{6} in a qualitative way,
but some quantitative differences in the values of the parameters of the
steady-state solution in the form of a stably propagating bipolar solitary
wave.


\subsubsection{3D generalization of the sine-Gordon equation}

We further simulated the propagation of the electromagnetic pulse within the
framework of the model represented by the truncated effective Eq.~\eqref{7},
for the same values of the system parameters and with the same initial
condition as before. It follows from the results of the calculations that the
pulse of model~\eqref{7} at long time scales also possesses the properties of
a breather---a bipolar solitary wave---with a periodically changing
``breathing" amplitude. Model~\eqref{8} with the corrected coefficient
$\sigma_\Sigma^2$ in front of the sine leads to a similar picture of the
evolution of the pulse field: after the transient stage (a short-term increase
in the degree of transverse field localization with a corresponding increase
in amplitude), the pulse reaches a stable propagation mode, qualitatively
similar to that in the framework of model~\eqref{5}.


\subsection{Comparison of the pulse dynamics in the framework of complete and
  reduced models}

To compare the behavior of the solutions obtained in the framework of
different models represented by Eqs.~\eqref{5}--\eqref{8}, we analyze the time
dependencies of the scaled amplitude of the maximum field intensity
$I_{\rm{max}}/I_0$, and the ratio of the transverse half-width of the pulse
$\lambda_\perp (\tau)$ to its initial value $\lambda_\perp (0)$ (see
Eq.~\eqref{19}).
\begin{figure}[tbp]
\centering
  \includegraphics[width=0.6\textwidth]{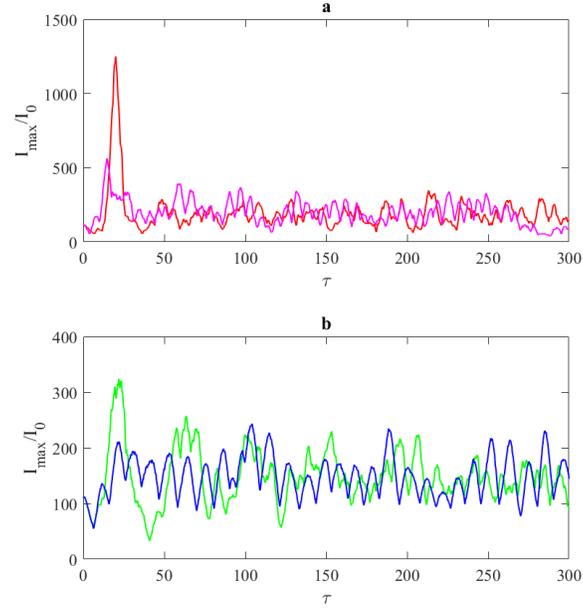}
  \caption{Dependency of the scaled maximum energy density of the pulse
    $I_{\rm{max}}/I_0$ on the dimensionless time $\tau$ along the $\zeta$-axis
    in the course of the electromagnetic pulse propagation in the full
    (Eq.~\eqref{5}) and truncated models represented by
    Eqs.~\eqref{6}--\eqref{8}: (a) red -- full effective Eq.~\eqref{5};
    magenta -- truncated model in the form of the three-dimensional
    generalization of the sine-Gordon Eq.~\eqref{8} containing the coefficient
    $\sigma_\Sigma$; (b) green -- truncated model in the form of the
    three-dimensional generalization of the double sine-Gordon Eq.~\eqref{6};
    blue -- truncated model in the form of the three-dimensional
    generalization of the sine-Gordon equation~\eqref{7}.}
  \label{fig5}
\end{figure}

\begin{figure}[tbp]
\centering
  \includegraphics[width=0.6\textwidth]{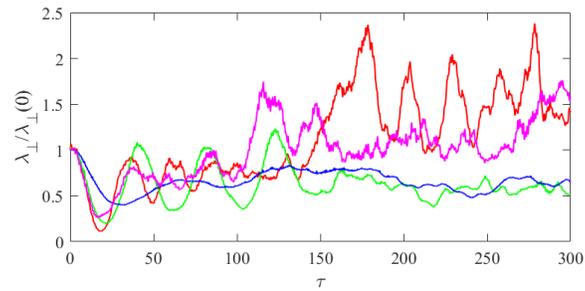}
  \caption{Dependency of the scaled transverse half-width of the pulse
    $\lambda_\perp/ \lambda_\perp (0)$ on the dimensionless time $\tau$ during
    the propagation of the electromagnetic pulse under the same conditions as
    in Fig.~\ref{fig5}: red -- full effective Eq.~\eqref{5}; green --
    truncated model in the form of a three-dimensional generalization of the
    double sine-Gordon equation~\eqref{6}; blue and magenta -- truncated model
    in the form of the three-dimensional generalization of the sine-Gordon
    equation~\eqref{7} and~\eqref{8}, respectively. }
  \label{fig6}
\end{figure}

Figures~\ref{fig5} and~\ref{fig6} show, respectively, the time dependencies of
the values of $I_{\rm{max}}/I_0$ and $\lambda_\perp /\lambda_\perp (0)$,
obtained as a result of solving the effective Eqs.~\eqref{5}--\eqref{8}
at identical values of the system parameters and the same initial condition
given by formulas~\eqref{20}--\eqref{25}. It can be seen from Fig.~\ref{fig5}
that a noticeable difference in the behavior of the solution of Eq.~\eqref{8},
with the modified coefficient in front of the sine, from the behavior of the
solutions within the remaining truncated models~\eqref{6} and~\eqref{7}, is a
more pronounced outgrowth of the field amplitude with the field focusing at
the initial stage, which is similar to the behavior of the pulse within the
framework of the model represented by the full effective Eq.~\eqref{5}. The
amplitude and transverse width of the pulse reach values close to the initial
values at the end of the transient process again, and the pulse dynamics
reaches the mode corresponding to the stable propagation of a breather, which
is the common result obtained within all the analyzed
models~\eqref{5}--\eqref{8}.  We also note that the truncated model in the
form of Eq.~\eqref{8} with the modified coefficient $\sigma_\Sigma$, in the
most accurate manner (with respect to the full Eq.~\eqref{5}) approximates the
variations of the transverse width of the electromagnetic pulse due to
diffraction.  This conclusion follows from the analysis of the curves
$\lambda_\perp (\tau)$ in Fig.~\ref{fig6}.

Thus, comparing the numerical simulation results shown in Figs.~\ref{fig5}
and~\ref{fig6}, it can be seen that the model represented by the
three-dimensional generalization of the sine-Gordon and the double sine-Gordon
equations can adequately approximate the dynamics of the pulse at long time
scales.  We note that the truncated model in the form of a generalization of
the sine-Gordon equation with the corrected coefficient calculated from the
empirical formula~\eqref{9}, is the simplest and satisfactory model for the
asymptotic analysis of the electromagnetic pulse dynamics at the long time
scales.


\section{Conclusions}


Key results of this work may be summarized as follows:
\begin{itemize}
\item[i)] An effective equation (see Eq.~\eqref{5}) is presented that
  determines the evolution of the electromagnetic field in an array of
  semiconductor carbon nanotubes, taking into account the limiting
  (three-dimensional) field localization.

\item[ii)] The evolution of an extremely short three-dimensional bipolar
  electromagnetic pulse in an array of nanotubes is analyzed by means of
  numerical simulations over a long time interval, which significantly exceeds
  the pulse durations, and still shorter than the relaxation time.  The
  possibility of stable propagation of a bipolar electromagnetic pulse with
  limiting (three-dimensional) field localization in an array of nanotubes at
  large time scales is confirmed.

\item[iii)] The possibility of approximating the effective Eq.~\eqref{5} by
  simpler models in the form of three-dimensional generalizations of the
  double sine-Gordon equation~\eqref{6} and the sine-Gordon
  equation~(\eqref{7}--\eqref{8}) is suggested and numerically verified.

\item[iv)] The dynamics of the pulse is compared in the framework of the
  models associated with the full effective Eq.~\eqref{5} and the truncated
  effective Eqs.~\eqref{6}--\eqref{8}. It is established that the truncated
  effective equations are adequate models for the description of the dynamics
  of a three-dimensional electromagnetic pulse in an array of carbon nanotubes
  over long times.
\end{itemize}

\section*{Appendix}

\subsection*{Possibilities of analytical consideration. Asymptotic analysis of the
  electromagnetic pulse dynamics}

The applicability of Eqs.~\eqref{6}--\eqref{8} proposed above opens the
possibility for further progress in the analytical study of the problem of the
dynamics of extremely short electromagnetic pulses in carbon
nanotubes. Consider the illustration of an analytical approach to the
asymptotic analysis of the field evolution over long times using the example
of the model described by Eq.~\eqref{7} (generalization to Eq.~\eqref{6} is
simple).

After re-normalization of time $\tau\to\sigma\tilde{\tau}$ and coordinates
$\xi\to\sigma\tilde{\xi}$, $\upsilon\to\sigma\tilde{\upsilon}$, and
$\zeta\to\sigma\tilde{\zeta}$, Eq.~\eqref{7} can be considered as the
Euler-Lagrange equation for the Lagrangian density,
\begin{equation}
  {\cal L} = \frac{1}{2}\left\{ \left( \frac{\partial\Psi}{\partial\tilde{\tau}} \right)^2 -
    \left( \frac{\partial\Psi}{\partial\tilde{\xi}} \right)^2 - \left( \frac{\partial\Psi}{\partial\tilde{\upsilon}} \right)^2
    - \left( \frac{\partial\Psi}{\partial\tilde{\zeta}} \right)^2\right\} + \cos\Psi -1.
  \label{A8}
\end{equation}
The change to the cylindrical coordinate system ($\tilde{\upsilon} = r
\cos\theta$, $\tilde{\xi} = r \sin\theta$), as well as the assumption that the
extremely short pulse preserves the cylindrical symmetry in the course of its
propagation ($d\Psi/ d\theta\rightarrow 0$) allows us to represent the
Lagrangian density~\eqref{A8} in the form
\begin{equation}
  {\cal L} = \frac{1}{2}\left\{ \left( \frac{\partial\Psi}{\partial\tilde{\tau}} \right)^2 -
    \left( \frac{\partial\Psi}{\partial r} \right)^2
    - \left( \frac{\partial\Psi}{\partial\tilde{\zeta}} \right)^2\right\} + \cos\Psi -1.
  \label{A9}
\end{equation}

Next, consider, according to the Whitham approach~\cite{55}, a solution in the
form of a $2\pi$--kink, in which the ``fast" and ``slow" variables are clearly
distinguished:
\begin{equation}
  \Psi = 4\arctan \left\{ \rho (\tilde{\tau} - \tilde{\zeta}) - \Phi\right\},
  \label{A10}
\end{equation}
where $\rho$ and $\Phi$ are the fast and slow variables, respectively. Passing
further to the coordinates of the light cone for $\tilde{\tau}$ and
$\tilde{\zeta}$, and also averaging over $(\tilde{\tau} - \tilde{\zeta})$, we
obtain the corresponding Lagrangian density:
\begin{equation}
  {\cal L} = \rho\frac{\partial\Phi}{\partial Z} + \rho \left( \frac{\partial\Phi}{\partial r}\right)^2
  + \frac{1}{\rho} - \frac{\pi^2}{3\rho^3} \left( \frac{\partial\rho}{\partial r}\right)^2,
  \label{A11}
\end{equation}
where $Z = \tilde{\tau} + \tilde{\zeta}$.

The set of equations for the quantities $\rho$ and $\Phi$ corresponding to the
Lagrangian density~\eqref{A11} has a hydrodynamic form~\cite{55}:
\begin{eqnarray}
  \frac{\partial\rho}{\partial Z} &+& \frac{\partial}{\partial r} 
  \left(\rho \frac{\partial\varphi}{\partial r}\right) = 0, \nonumber \\
  \frac{\partial\varphi}{\partial Z} &+&\frac{1}{2}\left( \frac{\partial\varphi}{\partial r}\right)^2 -
  \frac{2}{\rho^2} = \frac{\pi^2}{6\rho^3} \left( \frac{\partial^2\rho}{\partial r^2}\right) -
  \frac{3}{2\rho} \left( \frac{\partial\rho}{\partial r}\right)^2,
  \label{A12}
\end{eqnarray}
where the notation $\varphi = -2\Phi$ is introduced for convenience of
presentation. An analogy with the hydrodynamics of an ideal fluid arises when
the right-hand side of Eq.~\eqref{A12} is equated to zero. In this case, the
quantity $\rho$ has the meaning of density, the quantity $\varphi$ corresponds
to the potential of the velocity field, and the quantity $-2/\rho^2\equiv
H(\rho)$ plays the role of enthalpy~\cite{56}. Considering the relationship
between enthalpy and pressure $P$ in the form $H(\rho) = \int\rho^{-1} dP$, we
consider the stability of our solution with respect to long-wavelength
transverse perturbations of such a kind that the right-hand side of
Eq.~\eqref{A12} can be neglected. In this case, taking into account the
results from hydrodynamics, it is necessary to satisfy the condition $P > 0$
for the stability of the solution.  Thus, the hydrodynamic analogy for the
model described by Eq.~\eqref{7} allows one to establish the stability of its
solutions in the form of kinks over arbitrary (long) time scales. Development
and application of this analytical approach with respect to solitary waves of
other types, including breathers, is also of interest, and we consider it as a
subject for further research.

\section*{Funding}
  A. V. Zhukov and R. Bouffanais are financially supported by the SUTD-MIT
  International Design Centre (IDC). N. N. Rosanov acknowledges the support
  from the Russian Foundation for Basic Research, Grant 19-02-00312, and from
  the Foundation for the Support of Leading Universities of the Russian
  Federation (Grant 074-U01). M. B. Belonenko acknowledges support from the
  Russian Foundation for Fundamental Research. 
\section*{Acknowledgments}  
  E. G. Fedorov is grateful to
  Prof. Tom Shemesh for his generous support. B. A. Malomed appreciates
  hospitality of the School of Electrical and Electronic Engineering at the
  Nanyang Technological University (Singapore).

\section*{Disclosures}
The authors declare that there are no conflicts of interest related to this article.







\end{document}